\documentstyle[mprocl,psfig]{article}

\def\Journal#1#2#3#4{{#1} {\bf #2}, #3 (#4)}

\def\APJ{\em ApJ}

\def\PRL{\em Phys. Rev. Lett.}
\def\PRD{{\em Phys. Rev.} D}

\def\be{\begin{equation}}
\def\ee{\end{equation}}
\def\bea{\begin{eqnarray}}
\def\eea{\end{eqnarray}}

\begin{document}

\title{The Matter Plus Black Hole Problem in Axisymmetry}

\author{ S. R. Brandt, J. A. Font }

\address{Max-Plank-Institut f\"ur Gravitationsphysik (Albert-Einstein-Institut)\\
Schlaatzweg 1, 14473 Potsdam, Germany}


\maketitle\abstracts{
We present preliminary results in our long-term project of
studying the evolution of matter in a dynamical spacetime. 
To achieve this, we have developed a new code to evolve
axisymmetric initial data sets corresponding to a black hole surrounded by matter fields.
The code is based on the coupling of two previously existing codes.
The matter fields are evolved with
a 2D shock-capturing method which uses the characteristic information of the
GR hydro equations to build up a linearized Riemann solver.
The spacetime is evolved with a 2D ADM code designed to
evolve a wormhole in full general relativity.
An example of the kind of problems we are currently investigating
is the on axis collision of a star with a black hole.
}

\section{Introduction}
We report on progress in our project of coupling a general relativistic
hydrodynamical code with a code to solve the Einstein field equations in
axisymmetric spacetimes. This code will be an important tool
to study the evolution of matter in a dynamical spacetime.
One of our first applications will be the study of head-on star-black hole
collisions. This is an important step toward the simulation of more
complex scenarios, such as coalescing binaries, one of the most promising
sources of gravitational radiation
to be detected by LIGO and VIRGO interferometers.

\section{The Hydro Code}
The hydro part of the code makes use of modern high-resolution shock-capturing
schemes in order to handle discontinuities in the solution. 
It has the capability of
using different linearized Riemann solvers and different cell-reconstruction
procedures to accurately solve the Riemann problem at every cell interface.
Mathematically, this code relies on the knowledge of the characteristic fields
of the Jacobian matrices of the system of equations of general relativistic
hydrodynamics when written, explicitly, as a hyperbolic system of balance
laws. Further details about the mathematical structure of the equations
can be found in \cite{Betal}. The hydro code has been extensively described in
\cite{Fetal}$^{\!,\,}$\cite{Betal}. 

\section{The Black Hole Code}
The black hole code is the same as the one developed at NCSA\cite{brandt} and is based on
the standard ADM formalism\cite{YorkIVP}.  
The metric is evolved for the full set of Einstein equations using a 3+1 explicit leap-frog
scheme with centered differencing.
The shift variables are used to eliminate some off-diagonal metric terms.  Having a
black hole built into the spacetime avoids possible coordinate problems at the origin
of the spherical coordinate system, as there is an isometric surface at a finite radius.

This code is also capable of evolving spacetimes with angular momentum, and/or gravitational
waves in combination with the matter fields.  It also has a number of utilities built-in, e.g.
routines 
to extract the quasi-normal gravitational radiation modes, and to track the motion of
apparent and event horizons. 

\section{Coupling and Preliminary Results}
The addition of matter to the initial value problem is a straightforward application
of the York procedure.  There are two basic configurations which we have considered
thus far, a donut of matter distributed about the equator
and a Gaussian ball of matter on the axis.  This latter
configuration represents a three body collision (since we have only used equatorial
plane symmetry thus far), but we intend to generalize this in future work and
evolve two body collisions.

The codes were coupled by allowing each to alternatively update the variables.  First
the hydro code takes a step, treating the spacetime metric as fixed,
then the black hole code takes a step treating the matter fields as fixed.
We have shown that our results compare well with an independently written
1D code\cite{FM} that uses a hyperbolic formulation
of the Einstein equations\cite{BoMaForm}.
The codes agree at early times, confirming the correctness and accuracy
of both codes, but disagree slightly at late times. 

Our 2D code can run either with matter fields turned off, or with a fixed general
relativistic background, or with a dynamic evolving background which
does not react to the presence of matter.  
The code can, therefore, reproduce results of tests that each independent code
(the matter and metric codes) passed. In addition, we have successfully compared our
numerically evolved spacetimes against analytic steady state accretion results.
Specific details about the structure of the code and the equations being integrated
will be reported elsewhere.

As an example of what sort of problems this code can be used to study, consider
the collision of the black hole with two sharply peaked blobs of dust that fall
onto the black hole along the axis.  In Fig. 1 we see the collapse of the lapse.
We show the lapse
as a function of $\eta$, the logarithmic radial coordinate, along the axis 
at several different and progressively later times.
Since the dust is already sharply peaked when the evolution begins,
we expect a singularity to be in the
near future of the evolution, and since the spacetime is maximally sliced ($K_a{}^a = 0$)
the fact that the lapse is collapsing there confirms this.

In Fig. 2 we see the growth of the radial metric function with time, essentially due to the
tidal forces of the black hole, leading to the phenomenon known as ``grid stretching.''
Again, the plot shows the value of the function along
the axis through a sequence of time steps.
In addition to the spike from the black hole, we see two additional spikes growing on either
side of the dust ball as a result of the tidal forces it creates.  This growth in the
radial metric function normally proceeds without limit, and eventually crashes the evolution
code.  We would expect this problem to be exacerbated in the region surrounding the dust
ball as the peaks are narrower and harder to resolve.

This problem can be cured, however, simply by making the dust ball more diffuse.  In
Fig. 3 we see the result of such an evolution which proceeds stably until about 70M,
at which time the bulk of the gravitational wave packet has propagated passed the detector.
Also shown in the plot is a fit against the first two harmonics of the $\ell=2$ gravitational
wave mode.  Clearly this agrees quite closely with the wave one expects to see for a
black hole with a mass comparable to the dust ball plus black hole system.

\section{Future directions}
In the near future we plan to use the code for studying a number
of astrophysical applications, as well as extending our set of code tests.
We plan to compare the code with results from perturbation theory,
to make an exhaustive study of the parameter space
of the head-on collision of a star with a black hole (the {\it star}
will be modeled by both dust and a polytrope),
dust shells imploding onto a black hole, accretion disks,
and a fully dynamical spacetime version
of the relativistic Bondi-Hoyle accretion onto a 
moving black hole. Some of these projects are currently
under way.

\begin{tabular}{ccc}
\psfig{figure=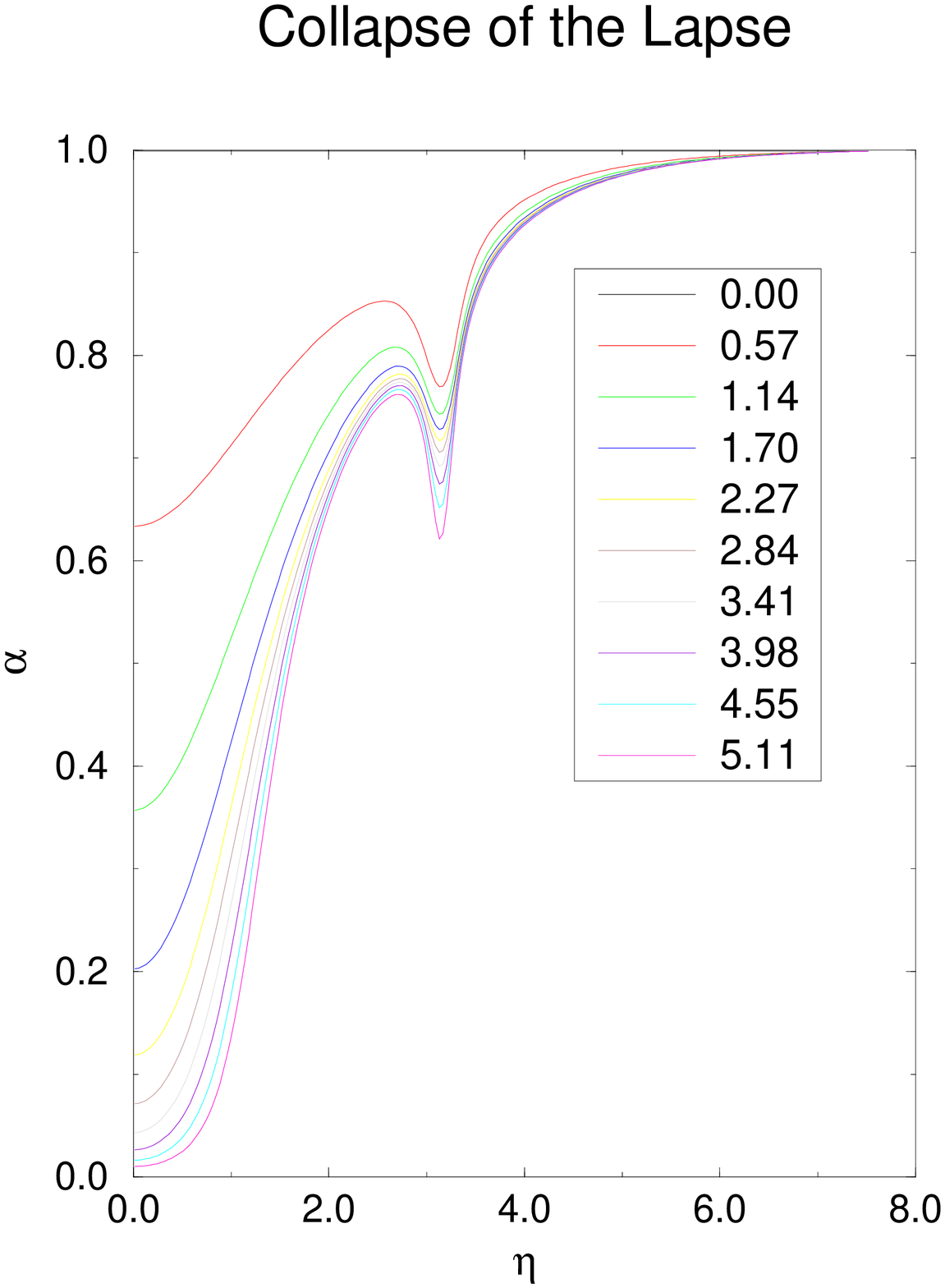,height=5.0cm,width=4.0cm} &
\psfig{figure=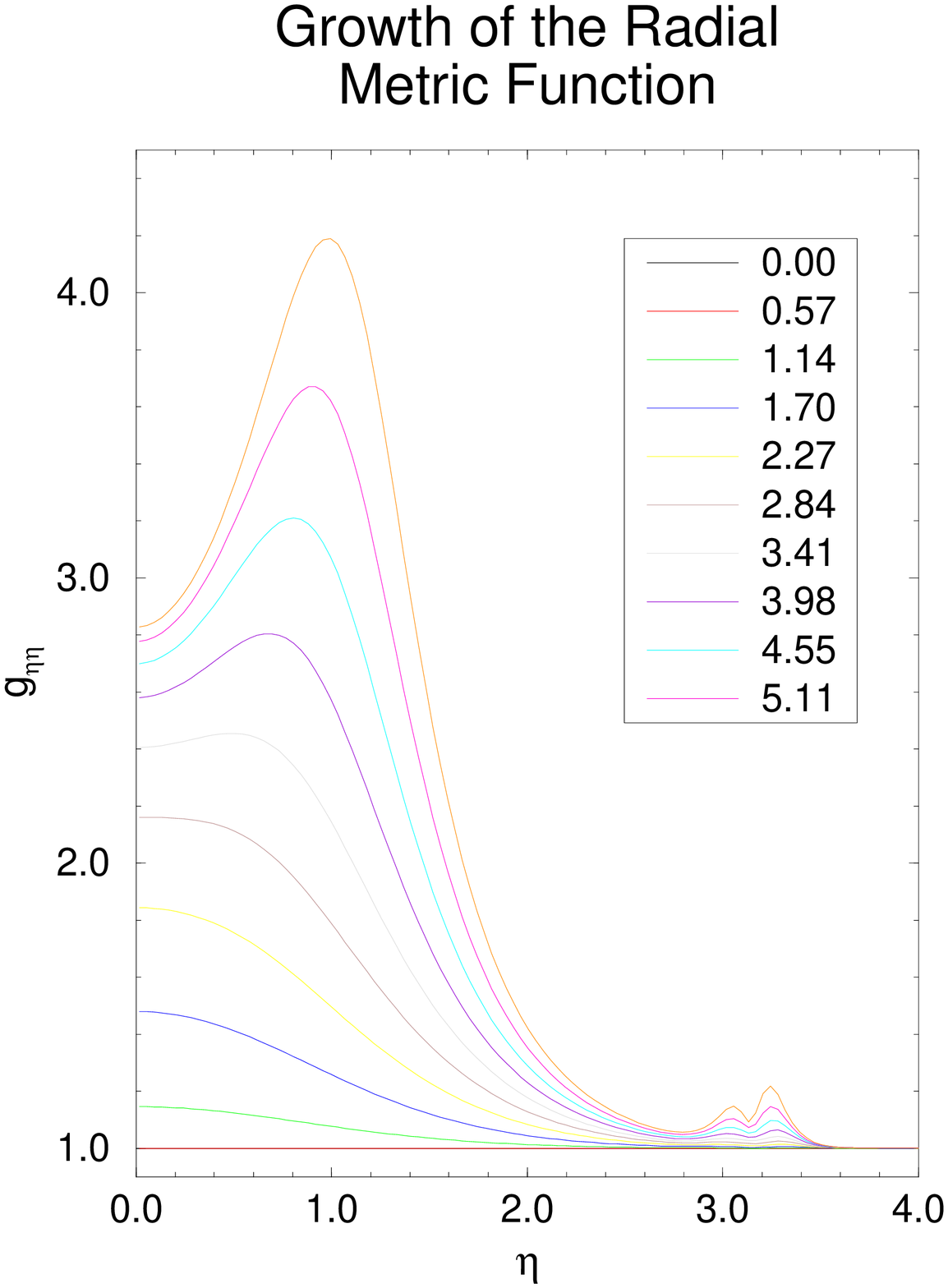,height=5.0cm,width=4.0cm} &
\psfig{file=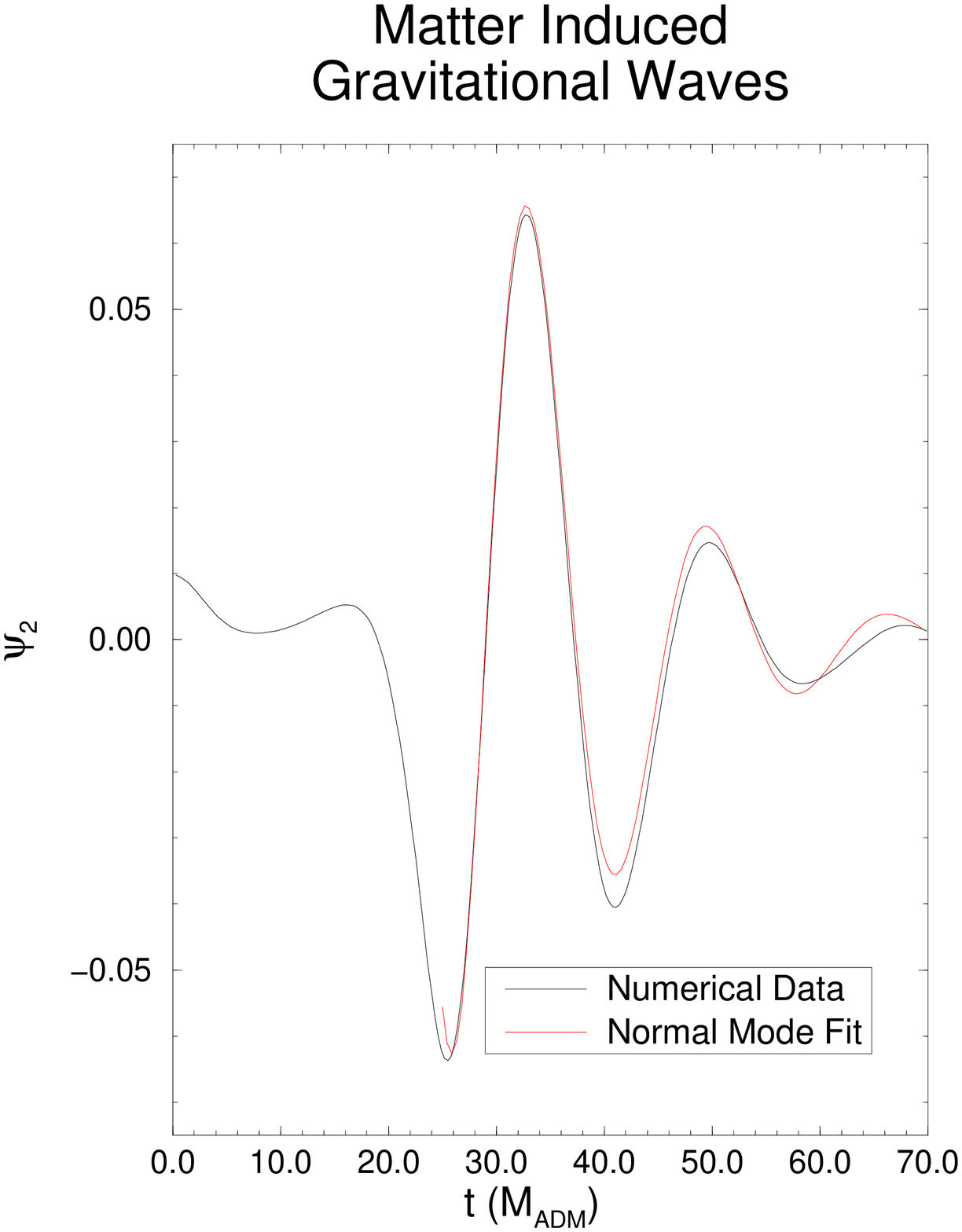,height=5.0cm,width=4.0cm} \\
\begin{minipage}[t]{4.0cm}
Figure 1: The collapse of the lapse as a function of time for
dust ball collapsing.
\end{minipage}&
\begin{minipage}[t]{4.0cm}
Figure 2: The growth of the radial metric function
along the axis for a dust ball colliding with a black hole.
\end{minipage}&
\begin{minipage}[t]{4.0cm}
Figure 3:  A gravitational wave for an initially diffuse dust ball
evolved to 70M.
\end{minipage}
\end{tabular}


\section*{References}
\begin{thebibliography}{99}

\bibitem{YorkIVP} James York in {\em Sources of Gravitational Radiation},
ed. L Smarr (Cambridge University Press, Cambridge, England, 1979).

\bibitem{Betal}F. Banyuls, J.A. Font,
  J.M$^{\underline{\mbox{a}}}$. Ib\'a\~{n}ez,
  J.M$^{\underline{\mbox{a}}}$. Mart\'{\i} and
  J.A. Miralles,
  \Journal{\APJ}{476}{221}{1997}.

\bibitem{brandt}S. Brandt and E. Seidel 
\Journal{\PRD}{52}{856}{1995};
{\bf 52}, {870}{(1995)};
{\bf 54}, {1403}{(1996)}

\bibitem{Fetal}J.A. Font, J.M$^{\underline{\mbox{a}}}$. Ib\'a\~nez,
A. Marquina and J.M$^{\underline{\mbox{a}}}$. Mart\'{\i}, 
 \Journal{AA}{282}{304}{1994}.

\bibitem{FM}J.A. Font and J. Mass\'o, this volume.

\bibitem{BoMaForm}C. Bona, J. Mass\'o, E. Seidel and
  J. Stela, \Journal{\PRL}{75}{600}{1995}.

\end{thebibliography}
\end{document}